\documentclass[useAMS,usenatbib]{mn2e}
\usepackage{graphicx,amssymb,color}
\usepackage[normalem]{ulem}

\title[Critical binary distance for white dwarf pollution]
{The critical binary star separation for a planetary system origin of white dwarf pollution}

\author[Veras, Xu, \& Rebassa-Mansergas]{
Dimitri Veras$^{1,2}$\thanks{E-mail: d.veras@warwick.ac.uk},
Siyi Xu$^{3}$,
Alberto Rebassa-Mansergas$^{4}$
\\
$^{1}$Centre for Exoplanets and Habitability, University of Warwick, Coventry CV4 7AL, UK
\\
$^{2}$Department of Physics, University of Warwick, Coventry CV4 7AL, UK
\\
$^{3}$European Southern Observatory, Karl-Schwarzschild-Stra{\ss}e 2, D-85748 Garching, Germany
\\
$^{4}$Departament de F\'{i}sica, Universitat Polit\'{e}cnica de Catalunya, c/Esteve Terrades 5, E-08860 Castelldefels, Spain
}
\pubyear{2017}

\begin{document}
\label{firstpage}
\pagerange{\pageref{firstpage}--\pageref{lastpage}}
\maketitle

\begin{abstract}
The atmospheres of between one quarter and one half of observed single white dwarfs in the Milky 
Way contain heavy element pollution from planetary debris. The pollution observed
in white dwarfs in binary star systems is, however, less clear, because companion star winds can generate
a stream of matter which is accreted by the white dwarf. Here we (i) discuss the necessity or lack
thereof of a major planet in order to pollute a white dwarf with orbiting minor planets in both single and binary systems, 
and (ii) determine the critical binary separation beyond which the accretion source is from a planetary system. 
We hence obtain user-friendly functions relating this distance
to the masses and radii of both stars, the companion wind, and the accretion rate onto the white dwarf, 
for a wide variety of published accretion prescriptions.
We find that for the majority of white dwarfs in known binaries, if pollution is detected, then that 
pollution should originate from planetary material.
\end{abstract}


\section{Introduction}

The unique properties of white dwarfs provide insights into planetary science which 
are currently unavailable in main sequence stars. The high density of white dwarfs, being
about $10^5$ as dense as the Earth, stratifies chemical elements in their atmospheres
\citep{schatzman1958,koester2009,altetal2010, koester2013} quickly, often on timescales of days
or weeks \citep[e.g. Fig. 1 of][]{wyaetal2014}.
Consequently, only hydrogen and helium should be observable in white dwarfs. 
The presence of elements heavier than He in a white dwarf's atmosphere, i.e. ``polluted" white 
dwarfs, would signify external accretion\footnote{An exception is for hot white dwarfs with 
cooling ages of under about 500 Myr, where some metals do not sink due to a process known as 
radiative levitation \citep[e.g.][]{koeetal2014}.}, and these have been observed.

In fact, single polluted white dwarfs have been known for a century 
\citep{vanmaanen1917,vanmaanen1919}. Only recently, however, have we fully understood the origin 
of the pollution. In addition, it has been found that between 25 and 50 per cent of single white 
dwarfs likely have photospheric heavy elements from accretion of planetary material 
\citep{zucetal2003,zucetal2010,koeetal2014}.
Recent studies, particularly from the Sloan Digital Sky Survey (\citealt*{dufetal2007}, 
\citealt*{kleetal2013}, \citealt*{genetal2015}, \citealt*{kepetal2015,kepetal2016}, and \citealt*{holetal2017}) 
have increased the total number of polluted white dwarfs to the thousands.

This polluted white dwarf population often contains observational signatures of Ca in the optical and Si 
in the ultraviolet. However, in over a dozen cases,
many more elements have been detected. In total, 20 different heavy elements have been found in white dwarf 
atmospheres 
\citep[e.g.][]{kleetal2010,kleetal2011,gaeetal2012,juretal2012,xuetal2013,
xuetal2014,wiletal2015,wiletal2016,xuetal2017}. Consequently, these white dwarf stars allow
us to probe the bulk chemical compositions of extrasolar planetesimals, providing for an unmatched 
window into internal structure \citep{juryou2014}. 

However, the overwhelming focus in white dwarf planetary science has been on single white dwarfs,
despite the fact that about three quarters of white dwarf--main sequence binaries are well-separated 
enough to bypass a common envelope phase \citep{wilkol2004}.
The reason for this focus is primarily because the planetary origin of white dwarf pollution becomes
uncertain in the presence of a binary companion; accretion from the wind of the secondary star is another possibility
\citep{debes2006}. \cite{zuc2014} found no significant difference in the occurrence of white dwarf pollution
in binaries if the binary separation is greater than 1000 au, and the discovery of white dwarf pollution from
a likely exo-Kuiper belt (volatile-rich) object \citep{xuetal2017} was in a binary system with a separation of a few thousand au. 

In this work, we focus on breaking this degeneracy from a theoretical perspective, and determine the
critical separation beyond which the planetary origin of white dwarf pollution holds in binary systems.
We provide explicit formulas which can be used for future discoveries to quickly characterise their planetary
nature, but also here discuss more broadly pollution in both single and binary systems. 
In Section 2, we reiterate the well-established reasons for the planetary nature of pollution in
single white dwarfs. Section 3 then provides an extensive discussion-based summary of why planets should be present 
in single polluted white dwarf systems. Section 4 extends these arguments to the binary case, 
and demonstrates that in these systems planets are not strictly necessary (but helpful).
In Section 5 we establish analytically criteria for assessing 
a planetary system origin of pollution in binary systems.
We apply our criteria to a sample of white dwarfs in binaries in Section 6,
and conclude in Section 7.

\section{Origin of white dwarf pollution}

Our strong statements about the utility of single white dwarf pollution as a probe of planetary
systems rely on the now-canonical assumption that the pollution does not arise from other
sources. The grounds for this justification are now well-trodden, but worth summarizing briefly
here due to the nature of this study.

An interstellar medium origin alone fails on chemical and kinematic grounds
\citep{aanetal1993,frietal2004,jura2006,kilred2007,faretal2010}. Even so, the 
planetary hypothesis was strengthened by the discovery of debris discs orbiting
white dwarfs
\citep[e.g.][]{zucbec1987,kiletal2005,reaetal2005,faretal2009,xujur2012,beretal2014},
whose tally is now over 40 \citep{farihi2016}.  All of the discs contain dust,
but some also contain gas. The discovery of gaseous disc components
\citep[e.g.][]{gaeetal2006,gaeetal2008,faretal2012,meletal2012,wiletal2014,manetal2016} 
demonstrated that dust and gas share the same location, which is
constrained to approximately $0.6-1.2R_{\odot}$.
This range demonstrates that the discs cannot have formed on the main sequence or giant
branch phases of stellar evolution: they must have been created during the white dwarf
phase. Combining that fact with the finding that every known disc orbits a polluted white
dwarf represents robust evidence for a planetary origin. Further, discs 
around some polluted white dwarfs may be too optically thin 
to be currently detectable \citep{bonetal2017}. Finally, and perhaps most
convincingly, one polluted white dwarf (WD 1145+017) which
also contains a debris disc was further discovered to host at least one minor
planet caught in the act of disintegrating
\citep{vanetal2015,aloetal2016,gaeetal2016,rapetal2016,redetal2016,xuetal2016,zhoetal2016,
croetal2017,garetal2017,guretal2017,haletal2017,kjuetal2017,veretal2017a}\footnote{\cite{faretal2017b} 
alternatively showed how the process of disintegration may be masked by suspended planetary dust
if the white dwarf is both magnetic and has a specific spin period.}.

\section{Planets in polluted single white dwarf systems}

As already outlined, evidence that white dwarf pollutants arise from planetary systems is overwhelming. However, do these systems need to contain actual planets, or is the existence of smaller bodies such as asteroids, moons, comets, pebbles and dust by themselves sufficient to explain the pollution?  At least six additional lines of evidence suggest that planets are necessary.

{\bf (i) mutual small body collisions:} Mutual interactions between debris in known belts (like the asteroid or Kuiper belt) predominantly result in collisional evolution rather than scattering towards the star, even for belts which are evolving during the giant branch phases of evolution \citep{bonwya2010}. Further, such outcomes might produce cascades which grind the debris into dust rather than orbitally perturbing it \citep{kenbro2017}.

{\bf (ii) comet / white dwarf collisions:} Direct collisions from exo-Oort cloud comets are rare, occurring at the rate of about one every $10^4$ yr \citep{alcetal1986} and sharply decreasing in frequency as a function of the white dwarf cooling age \citep{veretal2014b} in contrast to the observations \citep[Fig. 8 of][]{koeetal2014}. This rate is obtained from considering the excitation and depletion of an exo-Oort cloud due to mass loss, in combination with perturbations from stellar flybys and Galactic tides \citep{verwya2012,veretal2014c}. The rate might decrease if the giant branch star lost its mass anisotropically \citep{paralc1998,veretal2013a,caihey2017}.  These comets are anyway not guaranteed to remain intact upon approach, and instead might sublimate entirely at distances of up to several tens of au \citep{stoetal2015,broetal2017}.

{\bf (iii) sublimative perturbations:} An isolated volatile-rich body, such as an active asteroid (e.g. \citealt*{jewetal2015}), water-rich minor planet \citep{jurxu2010,jurxu2012,faretal2013a,radetal2015,malper2016,malper2017,genetal2017}, eccentric Kuiper belt object \citep{xuetal2017}, or exo-Oort cloud comet, can self-perturb its orbit due to stellar radiation-induced outgassing and sublimation.   However, for already highly eccentric bodies, these perturbations negligibly change the orbital pericentre \citep{veretal2015a}, meaning that outgassing effectively cannot push an object into a white dwarf Roche radius, and a perturbative force from a larger body (such as a planet) would be required.  Any body making a close approach to the Roche radius must already be highly eccentric ($e \gtrsim 0.98$; see \citealt*{veretal2014a}) in order for it to have avoided engulfment during the giant branch phase \citep{villiv2009,kunetal2011,musvil2012,adablo2013,norspi2013,lietal2014,viletal2014,staetal2016,galetal2017}.

{\bf (iv) long-distance transport:} Although dry asteroids are a much better chemical match to the abundance fractions seen in the atmospheres of white dwarfs \citep{juryou2014}, the asteroids must somehow reach the white dwarf Roche radius from a distance of at least 7-10 au.  Within this distance, but beyond the giant branch star's tidal engulfment radius, asteroids with radii in the range of 100 m - 10 km are subject to rotational fission from the radiative YORP effect \citep{veretal2014d}.  This effect arises from the vastly increased luminosity of the star during the giant branch phase (see figure 3 of \citealt*{veras2016a}), and may easily break apart asteroids which reside tens or hundreds of au away from the star.  Such distant intact asteroids would need a large (planet-mass) perturbing agent to reach the white dwarf. Liberated moons could perturb asteroids into the white dwarf \citep{payetal2016,payetal2017}, but the presence of moons of course implies the presence of planets\footnote{Moons liberated from their parent planets before the white dwarf phase and orbiting at a distance which is safe from giant branch engulfment could technically become planets or asteroids orbiting the white dwarf.}.

{\bf (v) dust migration:} The lack of observed cold dust \citep{juretal2007} indicates a lack of debris actively shrinking their orbits. The debris produced from YORP, or from mutual collisions, could pollute the white dwarf due to a gradual shrinking of their orbits due to Poynting-Robertson drag \citep{donetal2010,veretal2015b}, even at distances of up to 1000 au, at least for particles smaller than about 10 cm in size \citep{veretal2015c}.  If the debris was already on a highly eccentric ($e \sim 1$) orbit, then the orbital shrinking timescale for even 10 cm fragments at 1000 au is under about 1 Myr for a white dwarf cooling age of zero years, as is readily computed from Eq. 23 of \cite{veretal2015c}.  If, instead, the debris is on an initially near-circular orbit, then one can in principle derive a similar equation, or just solve Eqs. (111)-(112) of \cite{veretal2015b} numerically.  Doing so shows that fragments as large as 10 cm may be accreted at any point over a Hubble time.  However, the lack of detections of cold dust \citep{juretal2007} around polluted white dwarfs suggest that this process is not occurring.  Fragments larger than 10 cm are potentially influenced by the Yarkovsky effect, which can alter their orbits in a nontrivial manner \citep{veretal2015b}, but these should also have already produced observable signatures.

{\bf (vi) surviving planets:} Planets which are not engulfed by their eventual giant branch host star have to go somewhere.  Nearly every currently known exoplanet and Solar system planet orbiting beyond a couple au will survive into the white dwarf phase.  In fact, the relative fraction of known single white dwarfs which are polluted (25\%-50\%; \citealt*{zucetal2003,zucetal2010,koeetal2014}) is commensurate with the fraction of stars in the Milky Way galaxy that are currently thought to host planets \citep{casetal2012}.  All single planets will remain bound to their parent stars during giant branch mass loss, with perhaps just a handful of exceptions \citep{veretal2011,vertou2012,adaetal2013,veretal2013a}.  These planets then may interact with extant belts composed of smaller bodies to perturb them into the white dwarf to explain the observed accretion rates \citep{bonetal2011,debetal2012,frehan2014,antver2016,musetal2017,picetal2017}.  Multiple planets might undergo instability and gravitational scattering during the giant branch or white dwarf phases, but this process leaves at least one survivor in almost every case \citep{debsig2002,veretal2013b,voyetal2013,musetal2014,vergae2015,veretal2016,veras2016b}.

\section{Planets in polluted binary white dwarf systems}

We have just argued that planets should exist in polluted single white dwarf systems -- but what about
binaries containing a polluted white dwarf? If the binary is close enough, then the wind from the companion
could be either the dominant polluting source\footnote{One exception would be if both stars are white dwarfs \citep{heretal2014}.}
or represent an important contribution \citep{faretal2017a}. If the binary is sufficiently far apart, however, then
the pollution must arise from a planetary system. We will determine this critical separation in the subsequent
two sections. Here, however, we assume that the critical separation is surpassed and determine whether (major) planets
are necessary for pollution.

\subsection{Types of binary systems}

Consider first that in binary stellar systems, planets may orbit just one of the stars (sometimes known as ``S-type'') or orbit both stars in a circumbinary fashion (``P-type'').  Several tens of percent of all known exoplanetary systems are thought to be S-type binary systems\footnote{See e.g. the Exoplanets Data Explorer at http://exoplanets.org/}, whereas only about a couple dozen are securely P-type (e.g. \citealt*{sigetal2003} and \citealt*{doyetal2011}).  The high fraction of S-type systems demonstrates the importance of being able to distinguish a planetary origin of pollution in binary systems with a polluted white dwarf.  Three S-type systems (GJ 86, $\epsilon$ Ret, and HD 147513) actually do contain a white dwarf, although the planet orbits the other star \citep{butetal2006,desbar2007,faretal2011,faretal2013b}.

\subsection{P-type pollution}

Sufficiently wide binaries will not undergo Roche lobe overflow, nor experience a common envelope phase, both of which play crucial roles
in determining the survival of bodies orbiting both stars in a circumbinary fashion \citep{vertou2012,kosetal2016}. Surviving potentially
polluting bodies (e.g. asteroids) will likely be far away enough from both stars to necessitate perturbations from planets in order
to pollute the white dwarf\footnote{Assuming an isotropic belt of asteroids, both the white dwarf and companion star would be polluted in roughly
equal measure.}. The effect of radiation on smaller particles is not yet clear given the complexities of the orbit and the possibility of 
shadowing \citep[e.g.][]{rubincam2014}.

\subsection{S-type pollution}

Unlike for P-type pollution, S-type pollution -- assuming that the planetary system orbits the polluted star -- has been investigated by several studies, all using different architectural setups. 

\cite{bonver2015} demonstrated that for binary separations exceeding about $10^3$ au, Galactic tides will vary the eccentricity of the binary orbit significantly enough to potentially perturb a planetary system which is orbiting the white dwarf. They considered a single planet and a belt of planetesimals orbiting the white dwarf, all of which was coplanar with the binary companion. This mechanism and setup succeeded in contributing to metal pollution. Although they did not model a planet-less case, in principle the mechanism should also work there, based on the dynamics of the three-body problem.

Both \cite{hampor2016} and \cite{steetal2017} instead considered binary systems with a single other non-coplanar body. The mass of this body was sampled in a distribution that straddles traditional definitions of asteroid and planet (from 0.3 Mars masses to one Jupiter mass in \citealt*{hampor2016}, and for Eris and Neptune masses in \citealt*{steetal2017}). They showed that Lidov-Kozai oscillations (which feature oscillations of both eccentricity and inclination) can perturb the third body into the white dwarf, polluting it. Hence, these papers have presented another mechanism (like \citealt{bonver2015}) that does not necessarily require a planet in order to pollute the white dwarf in binary systems.

\cite{petmun2017} considered how an asteroid in a belt could remain stable to Lidov-Kozai perturbations {\it until} the white dwarf phase, {\it before} being thrust towards the white dwarf by such perturbations through the binary companion. They demonstrated that a planet which is eventually engulfed during the giant branch phase could stabilise in the main sequence stage this asteroid belt for long enough to represent a source of pollution. Although this method requires a planet, it does not necessarily do so along the white dwarf phase. 

These four studies demonstrate that a planet is not strictly necessarily in binary systems to generate pollution, although the lines are blurred depending on the definition of planet and the phase during which the planet exists.  Other considerations which limit the scope of planetary influence are how they are formed, and stability limits in post-main-sequence binary systems.  The observed binary separation in S-type planetary systems varies dramatically, from tens of au to thousands of au. However, this range is so large that placing constraints on the critical separation for which planet formation would be affected, disrupted or prevented is difficult.  Theoretical scenarios exist where planet formation may proceed in systems such as HD 196885, $\gamma$ Cephei, Gl 86, and HD 41004, where the binary separation is only about 20 au \citep[e.g.][]{rafikov2013}.  Nevertheless, the presence of a companion at several hundred au may easily affect the manner by and/or timescale over which a planet could form \citep{desbar2007}. As for multi-planet stability limits, \cite{veretal2017b} found that beyond 7 times the outer planet main sequence separation for circular binaries, a two-planet system will remain stable on the white dwarf phase. When instability does occur, the predominant outcome is ejection of at least one planet, although in up to a quarter of cases, one of the planets collides with the white dwarf (polluting it).

\section{Binary separation pollution criterion}

Now we only consider the accretion rate of the white dwarf due to its binary companion (ignoring all planetary bodies). Because the vast majority of known binaries containing white dwarfs also contain a main sequence star, henceforth we will assume that the white dwarf companion is a main sequence star. One-third of the known main sequence stars in white dwarf binaries are FGK stars \citep{paretal2016}, whereas the others are primarily M stars \citep{rebetal2013}.

\subsection{Main sequence stellar wind $\left[-\dot{M}_{\rm MS}(t)\right]$}

The critical separation beyond which a polluted white dwarf in a binary system can be assumed to have a planetary origin is significantly dependent on the stellar wind properties of the main sequence star. These properties are a function of that star's physical parameters, as well as age. Currently main sequence star winds are observationally poorly constrained, except for the Solar wind \citep{airusm2016}.  Consequently, we keep our treatment general in order to allow readers to insert their favoured values into the equations for future applications.  

The evolution of the Solar wind through time, $t$, is well-described by a power law with a exponent of $t^{-2.33 \pm 0.55}$ \citep{wooetal2005}, which is consistent with the $t^{-2}$ dependence reported in equation 4 of \cite{wooetal2002} and equation 9 of \cite{zenetal2010}. We adopt the functional form from \cite{zenetal2010} -- who specifically studied M stars -- and assume the main sequence star's mass loss proceeds as:

\begin{equation}
\dot{M}_{\rm MS}(t) = \dot{M}_{\rm MS}(t_{\rm i}) 
\left( \frac{t_{\rm i}}{t} \right)^x
\label{MSwind}
\end{equation}

\noindent{}where $\dot{M}_{\rm MS} < 0$ because the star is losing mass, $t_{\rm i}$ is the earliest time at which the formula is viable, $t \ge t_{\rm i}$, and $x>0$ is an arbitrary exponent.  \cite{zenetal2010} adopts $t_{\rm i} = 0.1$ Gyr, but in fact $t_{\rm i} = 0.7$ Gyr may be more realistic \citep{wooetal2005,airusm2016}.  For the Sun, $\dot{M}_{\rm MS}(t=t_{\rm i}) \approx -2 \times 10^{-11} M_{\odot}\,{\rm yr}^{-1}$, a value we adopt here. We also adopt $x=2.33$ and $t_{\rm i} = 0.1$ Gyr.

\subsection{Accretion rate onto white dwarf $\dot{M}_{\rm WD}(t)$}

The accretion rate onto the white dwarf from the wind can be described by a wide variety of not necessarily equivalent formulae (see references around Eq. 4.16 of \citealt*{veras2016a}), and is complicated by a number of factors \citep[see discussion after equation 2 of][]{debes2006}. The differences amount to alternative ways of treating gravitational focusing, and more specifically, the ``accretion radius'', under not necessarily mutually-exclusive models of Bondi-Hoyle accretion, spherical accretion and/or wind accretion. In order to highlight the sensitivity of the results depending on the prescription used, we adopt four significantly different published formulae from the post-main-sequence literature \citep{dunlis1998,huretal2002,jura2008,villiv2009}.

In order to place all formulae these on equal footing, we derive expressions for the density of the ambient medium by assuming that the stellar wind from the main sequence star emanates in a spherically symmetric manner and has a constant velocity (see Eq. 2 of \citealt*{hadjidemetriou1966}). Further, where the (unknown) wind speed needs to be specified explicitly, we adopt the approximate expression $(1/2)\sqrt{GM_{\rm MS}/R_{\rm MS}}$ from equation (9) of \cite{huretal2002} (justified by being proportional to the escape velocity of the main sequence star). The accretion rate onto the white dwarf is assumed to be an average over an orbit (so that the accretion rate is not a function of the mean anomaly) and that the mass loss is isotropic. If the mass loss is anisotropic, the equations become significantly more complex \citep{veretal2013a,doskal2016a,doskal2016b} and, for main sequence stars, is a largely unnecessary generalization.

We can also make simplifications to reduce the number of degrees of freedom in the equations and eliminate some of time dependencies. Consider

\begin{itemize}

\item {\bf separation vs. semimajor axis} The instantaneous separation $r(t)$ at some particular observed age $t=t_{\rm obs}$ 
is related to the orbital semimajor axis $a(t)$ and the orbital eccentricity $e(t)$ through

\begin{equation}
r(t_{\rm obs}) = 
\frac{a(t_{\rm obs})\left(1 - e(t_{\rm obs})^2\right)}
{1 + e(t_{\rm obs})\cos{\left[f(t_{\rm obs})\right]}}
\end{equation}

\noindent{}where $f(t_{\rm obs})$ is the true anomaly.  The observable, however, is the projected separation, $r_{\rm proj}(t_{\rm obs})$, which can range anywhere from $\left[a(t_{\rm obs})\left(1-e(t_{\rm obs})\right)\right]$ to $\left[a(t_{\rm obs})\left(1+e(t_{\rm obs})\right)\right]$, but is otherwise unknown. For practical purposes, we will make the approximation that 

\begin{equation}
r_{\rm proj}(t_{\rm obs}) \approx a(t_{\rm obs}).
\label{inst}
\end{equation}

\item {${\mathbf a(t)}, {\mathbf e(t)}$}: Despite the approximation in equation \ref{inst}, the semimajor axis $a(t)$ and eccentricity $e(t)$ of the mutual orbit may still change over the course of an orbit, and hence affect the mean accretion rate onto the white dwarf, due to the effect of Galactic tides \citep{heitre1986,brasser2001,fouchard2004,fouchard2006,vereva2013,veretal2014c}, with important consequences for white dwarf pollution \citep{bonver2015}.  However, the timescales for the change are many orders of magnitude longer than the orbital period, even for orbital periods of $10^7$ yrs.  Hence, we assume that $a(t)$ and $e(t)$ are fixed for our computations and are independent of time, and hence that
$r_{\rm proj}(t_{\rm obs}) \approx a(t_{\rm obs}) = a$. 

\item {${\mathbf M_{\rm MS}(t)}, {\mathbf M_{\rm WD}(t)}$}:  The value of $M_{\rm MS}(t)$ may be assumed to be independent of time, and the consequent effect on orbital expansion of the binary can be safely ignored: \cite{verwya2012} found that between now and the end of the Sun's main sequence lifetime, the semimajor axes of the Solar system planets will expand by at most about $0.055 \%$ due to mass loss. For similar reasons, we assume $M_{\rm WD}(t)$ also remains independent of time. 

\item {\bf ${\mathbf R_{\rm MS}(t)}, {\mathbf R_{\rm WD}(t)}$}: Unlike its mass, the radius of a main sequence star ($R_{\rm MS}(t)$) changes appreciably (by several tens of per cent) depending on the star's age. Hence, we treat $R_{\rm MS}(t)$ as time-dependent. Alternatively, the radius of a white dwarf ($R_{\rm WD}(t)$) is not assumed to change appreciably over at least one Hubble time, and so we consider $R_{\rm WD}$ fixed and time-independent.

\end{itemize}

\subsubsection{Hurley et al. (2002) formulation}

We start with the formulation of Eqs. (6-9) of \cite{huretal2002}, which is our fiducial prescription. 
Of our four chosen prescriptions, this prescription is the only one to specifically consider accretion 
in binary stellar systems\footnote{It is also the only one to assume the binary orbit is eccentric,
although this consideration does not usually vary the final result by more than an order of magnitude.}. These authors
provide the following mean accretion rate, which can be rewritten and expressed in our notation as

\[
\langle \dot{M}_{\rm WD}(t) \rangle^{({\rm H})} =
- 12\dot{M}_{\rm MS}(t) 
\left[ \frac{M_{\rm WD} R_{\rm MS}(t)}{a M_{\rm MS}} \right]^2
\]
\begin{equation}
\ \ \ \ \ \ \ \ \ \ \times 
\left[1 + \frac{4 R_{\rm MS}(t) \left[M_{\rm MS} + M_{\rm WD} \right] }{a M_{\rm MS}} 
\right]^{-3/2}
.
\label{bonhoy}
\end{equation}

\subsubsection{Jura (2008) formulation}

\cite{jura2008} instead considered accretion in a different context: the mass 
accreted by an asteroid due to stellar wind after the star has left the main sequence.
He assumed a circular orbit and no gravitational focusing. 
Applying the classical Bondi-Hoyle accretion formulation to our binary star 
system (transforming his asteroid into our white
dwarf and his giant branch star into our main sequence star) yields

\[
\langle \dot{M}_{\rm WD}(t) \rangle^{({\rm J})}
=
-\frac{1}{2} \dot{M}_{\rm MS}(t) \left(\frac{R_{\rm WD}}{a} \right)^2
\]

\begin{equation}
\ \ \ \ \ \ \ \ \ \ \ \ \times
\sqrt{\frac{R_{\rm MS}(t) \left(M_{\rm MS} + M_{\rm WD}\right)}{aM_{\rm MS}}  }
.
\label{jurorig}
\end{equation}

\subsubsection{Villaver \& Livio (2009) formulation}

The next two references address accretion onto a planet (rather than an asteroid) during the giant branch phases
of stellar evolution. \cite{villiv2009} used the same classical prescription as \cite{jura2008}, but 
replaced the physical radius of the accreting body with an ``accretion radius'' derived from the body's
orbital speed. Their equation (1) hence translates into our binary star setup as 

\[
\langle \dot{M}_{\rm WD}(t) \rangle^{({\rm VL})}
=
-2 \dot{M}_{\rm MS}(t) \sqrt{\frac{R_{\rm MS}(t)}{a}}
\]

\begin{equation}
\ \ \ \ \ \ \ \ \ \ \ \ \times
\left[\frac{M_{\rm WD}^2}{M_{\rm MS}^{1/2} \left(M_{\rm MS} + M_{\rm WD}\right)^{3/2}} \right]
.
\label{villivorig}
\end{equation}

\noindent{}The authors also adopted an alternative prescription when the accretion radius
is comparable to the radius of the accreted body. This case is not applicable here because
for white dwarfs disc material is accreted at distances of several hundred white dwarf radii.

\subsubsection{Duncan \& Lissauer (1998) formulation}

Like \cite{villiv2009}, \cite{dunlis1998} estimated
the impact rate on a planet due to the stellar wind through their equation (2).
However, \cite{dunlis1998} do not use the same accretion radius, and include
additional factors based on the wind speed and escape speed as well as the orbital
speed. After being translated into our setup, their equation reads

\[
\langle \dot{M}_{\rm WD}(t) \rangle^{({\rm DL})}
=
-\frac{1}{4} \dot{M}_{\rm MS}(t) \left( \frac{R_{\rm WD}}{a} \right)^2
\]

\[
\ \ \  \times
\sqrt{1 + 4 \left( \frac{R_{\rm MS}(t)}{a} \right) \left( \frac{M_{\rm MS} + M_{\rm WD}}{M_{\rm MS}} \right)}
\bigg[1 + 8a
\]

\[
\times \left(\frac{R_{\rm MS}(t)}{R_{\rm WD}}\right) 
\left( \frac{M_{\rm WD}}{aM_{\rm MS} + 4 R_{\rm MS}(t) \left(M_{\rm MS} + M_{\rm WD} \right)} \right)  \bigg]
.
\]

\begin{equation}
\label{dunlisorig}
\end{equation}

\subsection{Final formulae}

Some algebraic manipulation allow us to establish critical formulae for our various prescriptions in terms of the 
following four ratios {\it only}:

\begin{eqnarray}
\mu &\equiv& \frac{M_{\rm WD}}{M_{\rm MS}},
\label{eqmu}
\\
\gamma &\equiv& \frac{\langle \dot{M}_{\rm WD}(t) \rangle}{\dot{M}_{\rm MS}(t)},
\label{eqga}
\\
\alpha &\equiv& \frac{R_{\rm MS}(t)}{a},
\label{eqal}
\\
\delta &\equiv& \frac{R_{\rm WD}}{a}.
\label{eqde}
\end{eqnarray}

\noindent{}These ratios are related by the following dimensionless equations for, 
respectively, the prescriptions from \cite{huretal2002}, \cite{jura2008}, 
\cite{villiv2009} and \cite{dunlis1998} as

\begin{equation}
\gamma^{({\rm H})} = -12 \mu^2 \alpha^2 
\left[1 + 4 \alpha \left(1 + \mu\right) \right]^{-3/2},
\label{nodim}
\end{equation}

\begin{equation}
\gamma^{({\rm J})} = -\frac{1}{2} \delta^2 \sqrt{\alpha \left(1 + \mu\right)},
\label{nodimJ}
\end{equation}

\begin{equation}
\gamma^{({\rm VL})} = -2 \sqrt{\alpha} \mu^2 \left(1 + \mu\right)^{-3/2},
\label{nodimVL}
\end{equation}

\begin{equation}
\gamma^{({\rm DL})} = -\frac{\delta}{4} 
\left[ \frac{8 \alpha \mu}{\sqrt{1 + 4\alpha \left(1 + \mu\right)} }
      + \delta \sqrt{1 + 4\alpha \left(1 + \mu\right)} \right]
.
\label{nodimDL}
\end{equation}

\noindent{}In the limit of large separations, where $\alpha \ll 1$, we obtain $\gamma^{({\rm H})} \approx -12 \mu^2 \alpha^2$.

The differences in these prescriptions, which can be seen immediately from the functional dependencies, are significant. The value of $\mu$ is at most a few, and hence is of order unity. However, both $\alpha$ and $\delta$ may be arbitrarily small down to about $5 \times 10^{-2}$ au (see Section 5.4), and always $\delta \ll \alpha$. Therefore, in the expression for $\gamma^{({\rm DL})}$, the first term dominates over the second.  Also, because $\delta \ll 1$ for any stable binary setup, not including gravitational focusing (as for $\gamma^{({\rm J})}$) may severely underestimate the accretion rate. Further, when $\alpha \approx 1$, then $\gamma^{({\rm H})} \approx \gamma^{({\rm VL})}$. However, when $\alpha \ll 1$, then $\gamma^{({\rm H})} \ll \gamma^{({\rm VL})}$. The expressions for $\gamma^{({\rm H})}$ and $\gamma^{({\rm VL})}$ contain three degrees of freedom, whereas those for $\gamma^{({\rm J})}$ and $\gamma^{({\rm DL})}$ contain four.

\begin{figure}
\includegraphics[width=8.8cm]{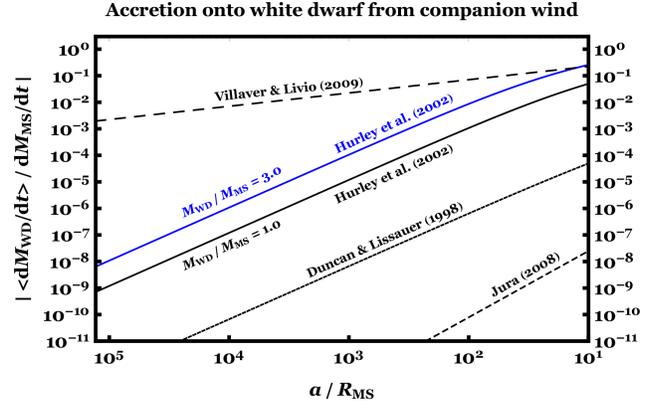}
\caption{
The amount of white dwarf pollution arising from the main sequence star wind
as a function of binary separation ($a$) in units of main sequence star radius ($R_{\rm MS}$).
The labels illustrate which of equations (\ref{nodim})-(\ref{nodimDL}) 
have been plotted for each line,
and the colours represent the given binary mass ratio.
The blue curve could represent, for example, a white dwarf--M dwarf binary.
}
\label{nodimfig}
\end{figure}

The three degrees of freedom in equations (\ref{nodim}) and (\ref{nodimVL}) allow us to plot their entire relevant solution set in Fig. \ref{nodimfig}. In addition, we plot equations (\ref{nodimJ}) and (\ref{nodimDL}) by assuming a physically plausible value of $\alpha = 300 \delta$. These equations and plots can be used to compute the amount of pollution from accretion of the main sequence wind. If at a given time $t = t_{\rm obs}$ one observes a higher rate of accretion onto the white dwarf than what is predicted from the equations, then likely the excess material arises from accretion of planetary debris. If instead one does not know $a$ (e.g. the binary system is not resolved) but observes accretion onto the white dwarf, then they could compute a critical separation beyond which pollution arises from planetary systems. 

\begin{figure}
\includegraphics[width=8.8cm]{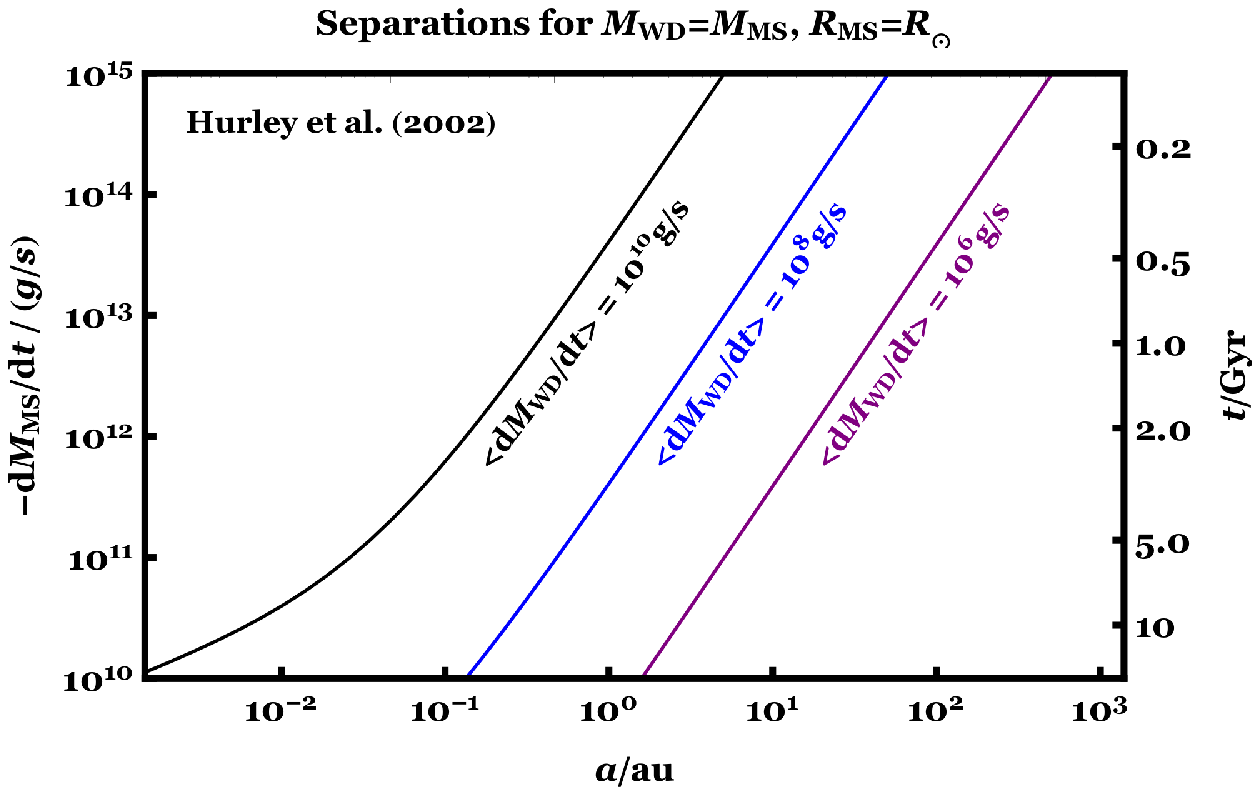}
\centerline{}
\includegraphics[width=8.8cm]{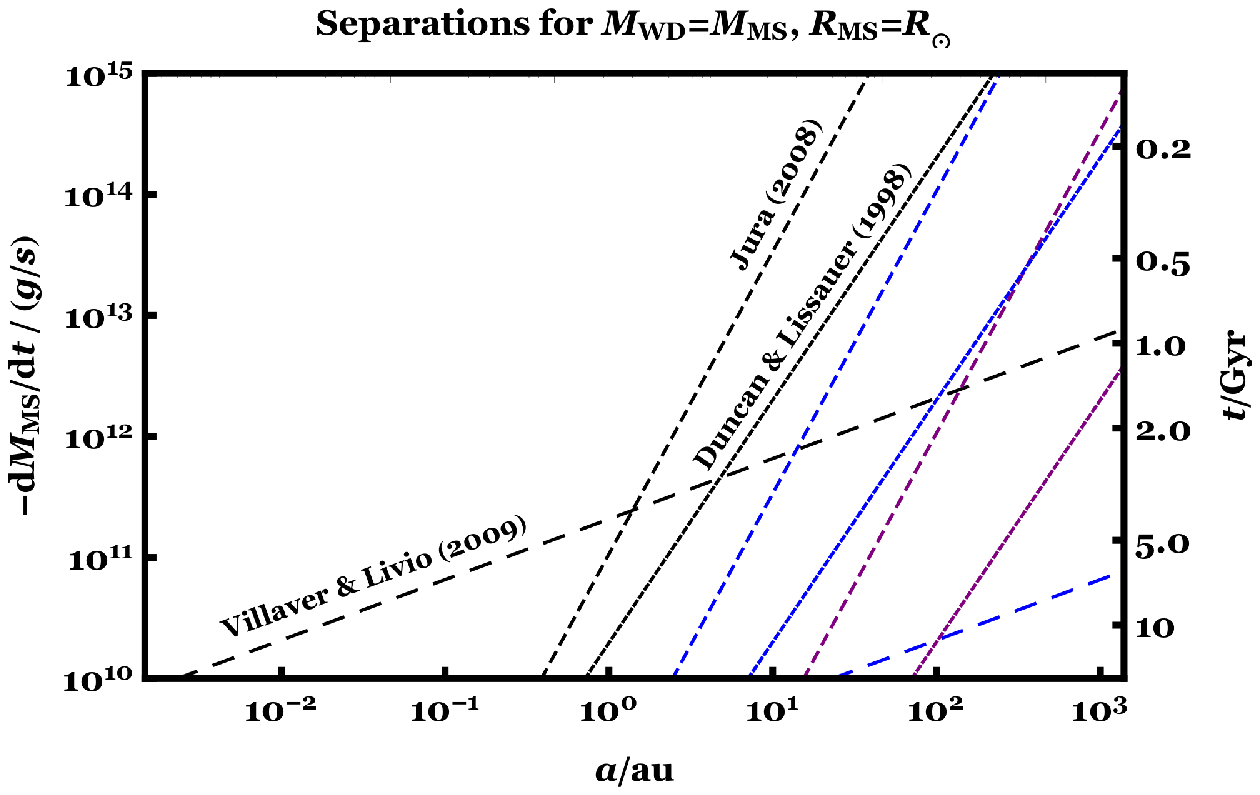}
\caption{
The relation between stellar wind mass loss rate (left $y$-axis),
main sequence stellar age (right $y$-axis), and binary separation
($x$-axis) for given values of the averaged white dwarf accretion rate,
assuming $M_{\rm WD} = M_{\rm MS}$ and $R_{\rm MS} = R_{\odot}$.
The upper plot is simply a special case
of equation (\ref{nodim}) and the Hurley et al. (2002) accretion curve
in Fig. \ref{nodimfig}, while
incorporating equation (\ref{MSwind}). The lower plot,
on exactly the same scale, illustrates how other accretion prescriptions 
(when extrapolated to stellar binaries) compare.
}
\label{critexample}
\end{figure}

Let us consider this latter case, and plot $a$ as a function of $\dot{M}_{\rm MS}$ assuming that 
$\langle \dot{M}_{\rm WD} \rangle \approx \left\lbrace 10^6, 10^8, 10^{10} \right\rbrace$~g/s,
$\mu = 1$ and $R_{\rm MS} = R_{\odot}$. The result, in Fig. \ref{critexample}, provides some concrete
examples of the scaled relations. The relevant curves in the lower figure also assume that $R_{\rm MS} = 300 R_{\rm WD}$.

\subsection{The minimum applicable separation}

Equations (\ref{nodim})-(\ref{nodimDL}) do not necessarily apply if 
both stars are close enough to 
be tidally interacting. In this case, the link between main sequence lifetime and stellar mass
loss rate would not necessarily hold.  Further, any binary pair which is close enough to
preclude planetary system material from orbiting just one of the stars would cloud
the interpretation of the origin of the pollution. Post-common-envelope systems
such as those containing WD 0710+741 (GD 448) and WD 2256+249 (GD 245, MS Peg)
\citep{zoretal2011,koeetal2014} would not be applicable here despite having enticing
measured accretion rates of some elements, because in both cases
the binary separation is $\lesssim 5 \times 10^{-2}$ au.

\section{Link with observations}

We can make comparisons with observations but with the caveat that only some subset of the six
variables in equations (\ref{eqmu})-(\ref{eqal}) are known for each system. 
In most cases, the white dwarf temperature  $T_{\rm WD}$, white dwarf surface gravity 
log $g$ (thus mass ${M}_{\rm WD}$ and $R_{\rm WD}$ if necessary), main sequence star 
mass $M_{\rm MS}$ and $a$ are measured.
The outstanding question is then, how does one compute the time-dependent
variables $\dot{M}_{\rm MS}(t)$ and $R_{\rm MS}(t)$?

\begin{figure*}
\includegraphics[width=14.8cm]{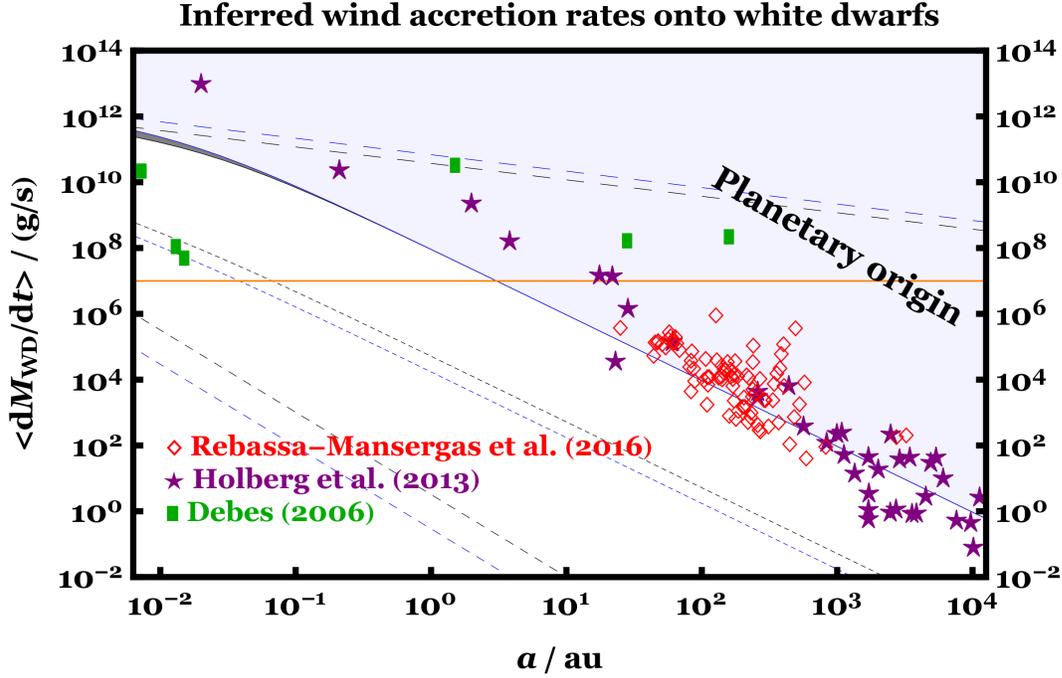}
\caption{
Inferred accretion rates onto white dwarfs from main sequence companion winds
from three samples: Holberg et al. (2013), Debes (2006), and Rebassa-Mansergas et al. (2016a). 
The solid and dashed pairs of curves correspond to the accretion prescriptions in Figs. 1-2 and approximate the boundaries
above which (in the shaded area for our fiducial prescription) newly observed polluted white dwarfs would likely
have remnant planetary systems, assuming $M_{\rm WD} = 0.6M_{\odot}$, $\dot{M}_{\rm MS} = -10^{12}$ g/s,
(black: $M_{\rm MS} = 0.6M_{\odot}$, $R_{\rm MS} = 0.6R_{\odot}$) and (blue: $M_{\rm MS} = 0.2M_{\odot}$, $R_{\rm MS} = 0.2R_{\odot}$). 
The orange horizontal curve roughly represents the current accretion rate detection limit of $10^7$ g/s. Consequently, if pollution is detected in a white dwarf which is separated from a binary companion by more than a few au, then very likely the pollution originates from a planetary system.
}
\label{holsample}
\end{figure*}

\subsection{Computing missing variables} \label{methvar}

If the white dwarf and the companion star were born at the same time (which might not be true for captured companions\footnote{This caveat is not particularly important for our considerations, as captured stars tend to have wider separations than the range which we are predominantly interested in.}), the age of the system can be calculated as the sum of the cooling age of the white dwarf $t_{\rm WD}^{({\rm cool})}$
plus the main sequence lifetime of that white dwarf $t_{\rm WD}^{({\rm MSlife})}$:

\begin{equation}
t_{\rm obs} = t_{\rm WD}^{({\rm cool})} + t_{\rm WD}^{({\rm MSlife})}
.
\end{equation}

\subsubsection{White dwarf cooling age $t_{\rm WD}^{({\rm cool})}$}

The cooling age may be related to the white dwarf's luminosity through equation 6
of \cite{bonwya2010} or equation 5 of \cite{veretal2015c} as

\begin{equation}
\frac{L_{\rm WD}}{L_{\odot}} \approx 3.26 \left( \frac{M_{\rm WD}}{0.6M_{\odot}}\right)
           \left(0.1 + \frac{t_{\rm WD}^{({\rm cool})}}{{\rm Myr}}  \right)^{-1.18}
\label{eqcool}
\end{equation}

\noindent{}assuming Solar metallicity. The luminosity can then be related to the
white dwarf effective temperature $T_{\rm WD}$ from

\begin{equation}
L_{\rm WD} = \sigma T_{\rm WD}^4 \left(4 \pi R_{\rm WD}^2 \right)
\label{eqLwd}
\end{equation}

\noindent{}where $\sigma$ is the Stefan-Boltzmann constant. The radius of the white 
dwarf $R_{\rm WD}$ can be calculated using the white dwarf mass radius 
relationship (equation 15 of \citealt*{verrap1988}):

\begin{equation}
\frac{R_{\rm WD}}{R_{\odot}} \approx 0.0127 \left(\frac{M_{\rm WD}}{M_{\odot}} \right)^{-1/3}
                                  \sqrt{1 - 0.607 \left(\frac{M_{\rm WD}}{M_{\odot}} \right)^{4/3}  } 
.
\label{eqRwd}
\end{equation}

\noindent{}Together, equations (\ref{eqcool}), (\ref{eqLwd}) and (\ref{eqRwd}) give us
$t_{\rm WD}^{({\rm cool})}$ given just $M_{\rm WD}$ and $T_{\rm WD}$.

\subsubsection{Main sequence age $t_{\rm WD}^{({\rm MSlife})}$}

In order to compute $t_{\rm WD}^{({\rm MSlife})}$, we need to employ an initial-to-final mass
relation and then a relation between main sequence mass and main sequence lifetime.
We do so by creating interpolating functions from the {\tt SSE} stellar evolution
code \citep{huretal2000}, assuming Solar metallicity, a Reimers mass loss coefficient
of $0.5$, and the default superwind prescription of \cite{vaswoo1993}. With these
assumptions, the only required input into the code is zero-age-main-sequence mass.
We created interpolating functions from the resulting grids of values.

\subsubsection{Main sequence radius $R_{\rm MS}$ and mass loss $\dot{M}_{\rm MS}$}

After finally obtaining $t_{\rm obs}$, we compute $\dot{M}_{\rm MS}(t_{\rm obs})$ from equation 
(\ref{MSwind}) and $R_{\rm MS}(t_{\rm obs})$ from a different interpolating function from the
same {\tt SSE} code.

\subsection{An ensemble of systems}

Having a prescription for obtaining all necessary variables, we 
can now consider how our criterion measures up to known white dwarf
binary samples. We consider three sets, from \cite{debes2006},
\cite{holetal2013} and \cite{rebetal2016a}, and plot values
from all in Fig. \ref{holsample}. The data is described below,
followed by a description of the critical boundary curves
on the plot.

\subsubsection{Holberg et al. (2013) data}

The sample from \cite{holetal2013} contains 
information about $M_{\rm WD}$, $M_{\rm MS}$, $a$ and $T_{\rm WD}$
from their Tables 1 and 2. From these values, we can estimate $\dot{M}_{\rm MS}(t)$
and $R_{\rm MS}(t)$ using the method in Section \ref{methvar}. Consequently, we can
derive $\dot{M}_{\rm WD}(t)$. If white dwarfs in these systems are observed to have values greater
than these inferred accretion rates, then our interpretation is that some of that 
pollution arises from planetary systems.

For this \cite{holetal2013} sample, we eliminate all systems which (i) lack a data point for at least any one
of $M_{\rm WD}$, $M_{\rm MS}$, $a$ and $T_{\rm WD}$ and (ii) do not conform to the method 
outlined in Section \ref{methvar}. These cuts leave us with 38 binary systems, which
we plot in Fig. \ref{holsample}. The points roughly fall along a diagonal line on the
log-log plot, one well approximated by the $\dot{M}_{\rm MS} = -10^{12}$ g/s assumption.
If observed accretion rates are higher than this line, then that
material is unlikely to arise from winds, and instead arise from planetary remnants.
The figure suggests that observing white dwarfs in binary systems with $a$ greater than a few au
for signatures of pollution may be worthwhile if typical accretion rates onto the white
dwarf exceed $\sim 10^7$ g/s.

\subsubsection{Debes (2006) data}

Notably, the \cite{debes2006} sample includes six
stars for which mass loss rates onto the white dwarfs have already been derived;
these estimations are not present in the other samples.
The rates are indicated by the filled rectangles in Fig. \ref{holsample}.
Three are close binaries, with separations of 0.0073 au (WD 0419-487),
0.013 au (WD 1026+002), and 0.015 au (WD 1213+528), and three are wide
binaries, with separations of 1.5 au (WD 0354+463), 28 au (WD 1049+103)
and 159 au (WD 1210+464). The wider binaries lie well within the
``planetary origin'' region of the plot.

\subsubsection{Rebassa-Mansergas et al. (2016a) data}

The final sample is the spectroscopic catalogue of white dwarf--M dwarf binaries from
the SDSS \citep{rebetal2016a}. With more than 3200 objects, this catalogue is
the largest and most homogeneous sample of white dwarf binaries to
date. From this sample we considered all white dwarf--main sequence binaries with available
effective temperatures, masses and radii for both the white dwarf and
the M dwarf components as well distance determinations. We note that
these parameters are obtained by applying the decomposition/fitting
routine outlined in \citet{rebetal2007} to the SDSS white dwarf--main sequence binary spectra. The
total age of the systems are calculated as the white dwarf cooling
time plus the white dwarf main sequence progenitor lifetime (see
\citealt{rebetal2016b} for details). Moreover, we only considered white dwarf--main sequence
binaries that are resolved in their SDSS images, as doing so allows
estimating the orbital separations as $a = \theta d$, where
$d$ is the distance to the binary and $\theta$ is the angular
separation of the two stars directly measured from the SDSS
images. What remains are the 85 white dwarf-main sequence binaries
plotted in Fig. \ref{holsample}.

\subsubsection{Curves on the figure}

The horizontal orange line represents a typical 
detectability threshold of $10^7$ g/s and each accretion 
prescription is given by two curves. As with the other figures,
the solid curves represent our fiducial prescription from \cite{huretal2002}.
The curves with the longest dashes are from \cite{villiv2009}, the curves
with medium dashes are from \cite{jura2008} and the curves with short
dashes are from \cite{dunlis1998}. For all pairs, $M_{\rm WD} = 0.6M_{\odot}$
and $\dot{M}_{\rm MS} = -10^{12}$ g/s (roughly the current solar value).
Within each pair, the black line corresponds to ($M_{\rm MS} = 0.6M_{\odot}$, 
$R_{\rm MS} = 0.6R_{\odot}$) and the blue line to 
($M_{\rm MS} = 0.2M_{\odot}$, $R_{\rm MS} = 0.2R_{\odot}$).
The difference in the two solid curves is small enough to be 
visible only at $a \sim 10^{-2}$ au. The difference in the curves
for the other pairs are more pronounced because those prescriptions
do not have the same cancellation due to the $\mu^2\alpha^2$ term.

For a given pair of curves if the observed 
accretion rates are higher than the curves, then that
material is unlikely to arise from winds, and instead 
arise from planetary remnants. Almost all observed
systems would therefore contain planetary remnants
if the \cite{jura2008} or \cite{dunlis1998} prescriptions
were used. Alternatively, wind accretion could explain
atmospheric metals in all but one or two of the observed
systems if the \cite{villiv2009} prescription is used.
We have shaded in the region
above the \cite{huretal2002} curves, the only prescription
given here specifically for binary star systems.

We do not believe that the \cite{jura2008}
nor \cite{villiv2009} prescriptions (the extreme cases
on Fig. \ref{holsample}) -- although certainly appropriate in
other contexts -- are appropriate
for our setup, for the following reasons.
\cite{jura2008} assumes accretion only if the target
(here the white dwarf) is hit directly. However,
the tidal, or Roche, radius of the white dwarf exceeds its 
physical radius by a factor of several hundred (at a 
distance comparable to $R_{\rm MS}$). On the other end, the accretion
radius adopted by \cite{villiv2009} is relatively large: on the
scale of au (about 1 au for an orbital velocity of 32.5 km/s
and about 46 au for an orbital velocity of 4.8 km/s).
This formulation arises from adopting orbital speed
and ignoring wind speed in \cite{villiv2009}'s definition of accretion radius. 
This choice amounts to including factors of $a$ instead of $R_{\rm MS}$
in the equation. Indeed, if we substituted $R_{\rm MS}$
for their accretion radius, then we would obtain 
$\gamma^{(\rm VL)} \propto \alpha^{5/2}$ rather than 
$\gamma^{(\rm VL)} \propto \alpha^{1/2}$.
The other prescriptions lie in-between these two extremes, 
and the more conservative and stellar-based one is from \cite{huretal2002}.

\section{Conclusion}

The overwhelming focus of planetary debris pollution studies has been single white dwarfs.
However, white dwarfs in binary systems may represent an unheralded supplement. Here,
we have argued that although the existence of major planets are a near-necessity in systems of
single polluted white dwarfs (Section 3), such planets need not exist in ``wide'' binary systems 
containing a polluted white dwarf (Section 4). We then determined just how wide
this separation needs to be in order for the origin of the pollution to be interpreted 
as arising from a remnant planetary system (but not necessarily one with a major planet). 

This critical separation can be given by one of
equations (\ref{nodim}, \ref{nodimJ}, \ref{nodimVL}, \ref{nodimDL}) depending on the 
reader's preference for accretion physics; we favour, in the context of binary white
dwarf / main sequence stars, the conservative but realistic 
prescription from equation (\ref{nodim}).
We hence find that the critical separation is at most a few au 
for observed accretion rates of $\sim 10^7$ g/s, the current
detectability threshold.
We hope our result will encourage observations of white dwarf atmospheres in binary systems,
which may provide unparallelled insights into binary-star planetary system evolution.

\section*{Acknowledgements}

We thank the referee for their very useful comments.
DV thanks ESO Garching for its hospitality during his visit in February 2016. He has received funding from the 
European Research Council under the European Union's Seventh Framework Programme (FP/2007-2013)/ERC Grant 
Agreement n. 320964 (WDTracer). ARM acknowledges financial support from MINECO grant AYA2014-59084-P and by the AGAUR.

\label{lastpage}
\end{document}